\documentclass[prd,a4paper,showpacs,superscriptaddress,twocolumn]{revtex4-1}
\usepackage{setspace}
\usepackage{subfigure}
\usepackage[utf8]{inputenc}
\usepackage{bm}
\usepackage{graphics}
\usepackage{amsmath, amssymb, amsfonts}
\usepackage{natbib}
\usepackage{hyperref}
\usepackage{color}
\hypersetup{colorlinks=true}
\usepackage[margin=0.53in]{geometry}
\usepackage{graphicx}
\usepackage{sidecap}
\usepackage{subfigure}
\usepackage[T1]{fontenc}
\usepackage{float}
\usepackage{dsfont}
\newcommand{\be}{\begin{equation}}
\newcommand{\ee}{\end{equation}}
\newcommand{\bse}{\begin{subequations}}
\newcommand{\ese}{\end{subequations}}
\newcommand{\ba}{\begin{eqnarray}}
\newcommand{\ea}{\end{eqnarray}}

\newcommand{\bea}{\begin{eqnarray}}
\newcommand{\eea}{\end{eqnarray}}
\usepackage{array}
\usepackage{lineno}
\usepackage{hyperref}
\usepackage{cleveref}
\newcolumntype{L}[1]{>{\raggedright\let\newline\\\arraybackslash\hspace{0pt}}m{#1}}
\newcolumntype{C}[1]{>{\centering\let\newline\\\arraybackslash\hspace{0pt}}m{#1}}
\newcolumntype{R}[1]{>{\raggedleft\let\newline\\\arraybackslash\hspace{0pt}}m{#1}}
\begin{document}
\title{Exact time dependence of causal correlations and nonequilibrium density matrices in holographic systems}
\author{Lata Kh Joshi}
\email{latamj@phy.iitb.ac.in}
\affiliation{Department of Physics, Indian Institute of Technology Bombay, Mumbai 400 076, India}
\affiliation{School of Mathematical Sciences, Queen Mary University of London, Mile End Road, London E1 4NS, United Kingdom}
\author{Ayan Mukhopadhyay}
\email{ayan@hep.itp.tuwien.ac.at}
\affiliation{Institut f\"ur Theoretische Physik, Technische Universit\"at Wien,
        Wiedner Hauptstrasse 8-10, A-1040 Vienna, Austria}
\affiliation{CERN, Theoretical Physics Department, 1211 Geneva 23, Switzerland}
\author{Florian Preis}
\email{fpreis@hep.itp.tuwien.ac.at}
\affiliation{Institut f\"ur Theoretische Physik, Technische Universit\"at Wien,
        Wiedner Hauptstrasse 8-10, A-1040 Vienna, Austria}
\author{Pichai Ramadevi}
\email{ramadevi@phy.iitb.ac.in}
\affiliation{Department of Physics, Indian Institute of Technology Bombay, Mumbai 400 076, India}
\begin{abstract}
We present the first exact calculations of the time dependence of causal correlations in driven nonequilibrium states in $(2+1)$-dimensional systems using holography. Comparing exact results with those obtained from simple prototype geometries that are parametrized only by a time dependent temperature, we find that the universal slowly varying features are controlled just by the pump duration and the initial and final temperatures only. We provide numerical evidence that the locations of the event and apparent horizons in the dual geometries can be deduced from the nonequilibrium causal correlations without any prior knowledge of the dual gravity theory.
\end{abstract}

\maketitle
\section{\textbf{Introduction}}Hydrodynamics gives us a general understanding of how expectation values of local operators e.g. the energy-momentum tensor and conserved currents in many-body systems thermalize. A similar general understanding of nonequilibrium time evolution of correlation functions is still elusive. This is particularly hard for the case of the unequal time correlation functions in which case nonperturbative techniques are essential even at weak coupling \cite{Berges:2002wr,Berges:2004yj,Calabrese:2006rx}. The chief objective of this work is to perform the first \textit{exact} calculations of the time dependence of the causal (a.k.a. retarded) correlations in a $(2+1)-$D many-body system in states transitioning from an initial thermal equilibrium to another driven by a \textit{homogeneous} energy injection from a source (a.k.a. pump). To that end we apply the holographic correspondence \cite{Hartnoll:2016apf} that maps strongly interacting many-body systems to classical theories of gravity with a few dynamical fields in one higher dimension \footnote{Although such holographic studies have been performed before, either the background geometry dual to the driven nonequilibrium state was approximated with simple analytic forms and/or the correlation function was calculated using geodesic-type approximations (references mentioned later). In the probe approximation, the geometry reduce to a simple AdS-Vaidya prototype and exact calculations of the nonequilibrium retarded correlator were performed by us in this context in our previous work.}. Furthermore, since the time dependent causal correlation function can be known experimentally via techniques such as solid-state pump-probe spectroscopy \cite{Segre:2001aa}, it is desirable to establish a general theory of its thermalization.

Our computations reveal that at least in the regimes where the time duration of energy injection $t_p$ is small compared to the initial thermal scattering time $T^{-1}_{\rm in}$, the time dependence of causal correlation functions in holographic systems has well-defined universal features. Namely, these features can be reproduced with better than $\mathcal{O}(t_p T_{\rm in})$ accuracy by a simple prototype gravitational geometry that can be constructed using \textit{only} (i) the experimentally controlled $t_p$ and (ii) the initial and final temperatures. Thus, the universal features of thermalization can be understood without detailed knowledge of microscopic dynamics whose only role here is to determine the final temperature of the system given the duration and amplitude of the external source. This indicates that many features of thermalization of the causal correlations are controlled via simple parameters analogous to the Reynolds number for hydrodynamic flows.

This result is tied to the second motivation of our work which is to get a deeper understanding of the holographic duality itself. One of the fundamental questions is, if a large-$N$ quantum system is holographic, can the dual classical gravity theory be constructed directly from the observables themselves? We can rephrase the question in this form: having measured the time-dependent expectation values of local operators and their correlation functions in a given nonequilibrium state, can we construct the dual gravitational geometry or at least know some of its defining features \textit{without} prior knowledge of the dual classical gravity equations?

We will demonstrate here that the universal features of the time-dependent causal correlations reveal the \textit{exact} location of the event horizon in the dual geometry. Furthermore, if we combine both universal and nonuniversal features of time-dependent causal correlations, we can extract the \textit{exact} location of the apparent horizon in an appropriate bulk coordinate system. Thus \textit{without} prior knowledge of the dual classical gravity equations, we can extract the locations of the event and apparent horizons of the dual geometries from measurements of causal correlations. Although knowing the location of the horizons in the dual geometries will not be sufficient for deducing the dual holographic classical gravity description, it will certainly be able to constrain the possibilities. Furthermore, it will constitute a significant step in understanding how the dual classical gravity can be decoded from the observables.

We can find indications regarding how we can unravel the enormous complexity of nonequilibrium states by understanding first simple density matrix approximations characterized only by an appropriately defined time-dependent temperature $T(t)$ in holographic theories. These density matrices will be dual to anti-de Sitter--Vaidya (AdSV) geometries with appropriately defined mass functions $M(v)$ where $v$ is an affine parameter along a null congruence. The dual density matrices/AdSV geometries are \textit{not} to be understood as solutions of the underlying microscopic dynamics/dual classical gravity equations. The AdSV prototype geometries most of which will be novel constructions do not approximate the exact nonequilibrium geometry in our pumping regimes. Nevertheless, we will learn a lot about the exact geometry from these prototypes depending on which (universal/specialized) aspects of the nonequilibrium retarded correlator they can approximate.

One particular AdSV prototype will require only the knowledge of the duration of the pump, and the initial and the final temperatures for its construction, and will be able to reproduce the features of the time-dependent retarded correlation function when the probe time is not within the pumping duration. Since the construction of this prototype involves no detailed knowledge of the underlying microscopic dynamics or the dual gravity theory, we claim these features of the nonequilibrium retarded correlation function that are reproduced within $\mathcal{O}(t_{p} T_{\rm in})$ accuracy by this prototype are universal and that they are only controlled by the mentioned parameters. However, this AdS-Vaidya geometry will not be able to reproduce any one-point function (energy density/pressure) even close to the same order of accuracy. Based on numerical results of how event horizons respond to external driving forces at the boundary, we will be able to intuitively explain in Sections \ref{exact} and \ref{horizons} why the nonequilibrium retarded correlation function is such a \textit{special} observable with features that can be reproduced from very simplistic AdS-Vaidya prototypes.

By construction, our other AdSV prototype geometries will reproduce the time dependence of at most one chosen one-point function (energy-density/pressure) or the location of either the event or the apparent horizon in the dual geometry exactly, but will fail to do so for all the other ones within $\mathcal{O}(t_{p} T_{\rm in})$ accuracy. Since the AdS-Vaidya geometries depend essentially on a function of one variable, namely $M(v)$, it can at best reproduce one chosen time-dependent function. However, we will provide numerical evidence that these prototype geometries will be able to approximate the universal or specialized features of the retarded correlation function with even better than $\mathcal{O}(t_{p} T_{\rm in})$ accuracy. This is nontrivial given that  no AdS-Vaidya construction can be designed to reproduce these features exactly  unlike the time dependence of a one-point function in the case of a typical pumping  protocol. Due to the lack of time-translation invariance, the nonequilibrium correlator depends on the probe time ($t'$) and the observation time ($t$) in a nontrivial manner, and not simply on $t - t'$ as in equilibrium. It is generically not possible to fit a function of two variables accurately by choosing a function of one variable. Therefore, it will be indeed a nontrivial result that the AdS-Vaidya geometries will be able to approximate some features of the  retarded correlation function in the far-from-equilibrium regime with better than $\mathcal{O}(t_{p} T_{\rm in})$ accuracy.

We will be able to provide numerical evidence that the AdS-Vaidya geometries which give best approximations to the exact retarded correlation function in specific domains of the probe and observation times will be able to reveal the locations of the event/apparent horizons as well. Furthermore, we also find that these AdSV constructions which reproduce the exact locations of the event and apparent horizons approximate the exact time-dependent pressure and energy density respectively to a remarkable accuracy even within the pumping duration although they are not designed (or expected) to perform such approximations. However, since one can design AdSVs which reproduce the energy density and the pressure exactly, the approximation of the retarded correlation function will be more crucial in deducing the mentioned AdSVs which reveal the apparent and event horizon dynamics in the dual classical gravity theory.


The plan of the paper is as follows. In Sec. II, we will describe the numerical construction of the classical geometries dual to the driven nonequilibrium state. In Sec. III, we will review our previously developed method of calculating the nonequilibrium retarded correlator and discuss the implementation. In Sec. IV, we will present the exact results, and divide the features into various categories, in particular depending on whether they can be reproduced by AdS-Vaidya prototypes within the desired accuracy and whether they depend on the pumping protocol. In Sec. 5, we will discuss the construction of the various AdS-Vaidya prototypes. In Sec. VI, we will use these prototypes to establish the universality of some features of the nonequilibrium retarded correlator, and show how from the latter we can deduce the locations of the horizons. Finally, in Sec. VII we will conclude with an outlook. The Appendices will provide supplementary details.

\section{\textbf{Driven nonequilibrium holographic states}}\label{drive} We consider a generic nonequilibrium state in a $(2+1)$-dimensional large-$N$ conformal field theory (CFT) driven by the Hamiltonian $H(t) = H_{\rm CFT} + H_{\rm pump}(t)$ from one thermal equilibrium to another with initial and final temperatures $T_{\rm in}$ and $T_{\rm f}$ respectively. Here, $H_{\rm pump}(t)$ represents energy injection from a \textit{homogeneous} external source $f(t)$ coupling to a scalar operator $O$ of the CFT, i.e. $H_{\rm pump}(t) = \int{\rm d}^2 \mathbf{x}\, f(t) O(\mathbf{x})$. For our specific construction, $O$ is assumed to have scaling dimension $\Delta = 2$ (like the electronic density operator at weak coupling). The external source $f(t)$ (which has the dimension of energy) is assumed to have a Gaussian profile:
\begin{equation}\label{eq:pump}
f(t) = E_{\rm max} e^{-t^2/2\sigma^2}.
\end{equation}
The effective duration of the pumping is thus $\vert t \vert  < t_{\rm p}/2$ with $t_{\rm p} \sim 6\sigma$. The final temperature $T_{\rm f}$ will be determined by the microscopic dynamics as functions of $T_{\rm in}$, $E_{\rm max}$ and $t_{\rm p}$. Since, the underlying microscopic theory is conformal, we choose units of measurement where $T_{\rm in} = 1$. We focus on the case $E_{\rm max}\sim T_{\rm in}$ and $t_{\rm p}T_{\rm in} \ll 1$ arguing later why our results will be independent of the specific choice of $f(t)$ as long as these conditions are satisfied.

Holographically, such a driven state is represented  by a $(3+1)$-dimensional asymptotically AdS metric $G_{MN}$, and a scalar field $\Phi$ (dual to the operator $O$) with mass given by $m^2 = -2/l^2$ and minimally coupled to Einstein gravity with cosmological constant $\Lambda = -3/l^2$ where $l$ denotes the radius of AdS space. It is convenient to choose coordinates where $G_{MN}$ and $\Phi$ take the form
\begin{eqnarray}\label{bmetric}
{\rm d}s^2 &=& \frac{l^2}{r^2}\left(-2{\rm d}r{\rm d}v - A(r,v) {\rm d}v^2\right) + S^2(r,v)({\rm d}x^2 + {\rm d}y^2),\nonumber\\
\Phi &=& \Phi(r,v),
\end{eqnarray}
where we have imposed homogeneity in the field-theory spatial coordinates $x$ and $y$. At the boundary $r=0$, the bulk coordinate $v$ is identified with the field-theory time coordinate $t$.

To achieve a unique gravitational solution, we need to provide (i) initial conditions for $S(r)$, $\Phi(r)$ and $\lim_{r\rightarrow 0}(1/6)\partial_r^3A(r) = a_{3\rm in}$ at $v =v_{\rm in}$ in the far past, and (ii) the boundary conditions $\lim_{r\rightarrow 0}A(r,v) = a_0(v)$, $\lim_{r\rightarrow 0}rS(r,v) = s_0(v)$ and $\lim_{r\rightarrow 0} r^{-1}\Phi(r,v) = \Phi_0(v)$. By the holographic dictionary $a_0(v) = s_0(v) =1$, since the dual system lives on flat Minkowski metric, and $\Phi_0(v)$ is identified with the source of the dual operator $f(v)$. Our initial state is thermal, and therefore the initial conditions are set via a black hole geometry with mass $M_{\rm in}$ so that $S(r, v_{\rm in}) = 1/r$, $a_{3\rm in} = -M_{\rm in} = -(4\pi/3) l^2 T_{\rm in}^3$ and $\Phi(r, v_{\rm in}) = 0$. The gravitational solution is obtained numerically via the method of characteristics \cite{Chesler:2008hg,Chesler:2013lia} as described in Appendix \ref{appen1}
. From this solution, we can extract the energy density $\langle t_{00} (t)\rangle$, the pressure $\langle t_{xx} (t)\rangle = \langle t_{yy} (t)\rangle$ and the expectation value $\langle O (t)\rangle$  (see Appendix \ref{appen2}
 for more details) in the dual driven state via the holographic renormalization procedure \cite{Skenderis:2002wp}.

\section{\textbf{Time-dependent causal correlations}}\label{prescription} Linear response theory tells us that if we perturb the time evolution of the nonequilibrium state driven by the pump via a probe perturbation $\Delta H = \gamma\int {\rm d}^2\mathbf{x} \tilde{f}(t,\mathbf{x})\tilde{O}(\mathbf{x})$, where $\tilde{O}$ is an operator which is the same as or different from $O$ to which the pump couples, then the time-dependent expectation value of $\langle\tilde{O}(\mathbf{k}) \rangle$ is given by:
\begin{equation}\label{causal}
\delta\langle\tilde{O}(\mathbf{k}) \rangle(t) = \gamma  \int {\rm d} t' G_R^{\tilde{O}\tilde{O}}(t, t',\mathbf{k}) \tilde{f}(t', \mathbf{k}) + \mathcal{O}(\gamma^2).
\end{equation}
Above $G_R^{\tilde{O}\tilde{O}}(t, t',\mathbf{k})$ is the causal correlation function (we have used the spatial homogeneity of the pump to Fourier transform the spatial $\mathbf{x}-\mathbf{x'}$ dependence). In the Heisenberg picture $G_R$ takes the form:
\begin{equation}
G_R^{\tilde{O}\tilde{O}}(t, t',\mathbf{k}) = -i \theta(t-t'){\rm Tr}(\rho_{\rm in} [\tilde{O}(t,\mathbf{k}),\tilde{O}(t',\mathbf{k})])
\end{equation}
with $\rho_{\rm in}$ being the initial thermal density matrix at temperature $T_{\rm in}$. Note that the time evolution operator $U(t,t') = T\exp \{-i\int {\rm d}t'' [H_{\rm CFT}+H_{\rm pump}(t'')]\}$ implicit in the definition above includes the pump, and therefore $G_R^{\tilde{O}\tilde{O}}(t, t',\mathbf{k})$ is not simply a function of $t-t'$ except in the far past and future when the pumping ceases and the states thermalize. Here, we will consider all possible cases in which the probe source $\tilde{f}(t)$ is turned on before, during or after the pump. For the sake of simplicity, we will consider $\tilde{O}$ to be an operator of scaling dimension $\Delta = 2$ too.

In Ref. \cite{Banerjee:2016ray} (see Refs. \cite{Skenderis:2008dg,Banerjee:2012uq,Balasubramanian:2012tu,Mukhopadhyay:2012hv,Skenderis:2008dg,Keranen:2014lna} for earlier related works), a holographic prescription has been developed for obtaining $G_R^{\tilde{O}\tilde{O}}(t, t',\mathbf{k})$ via a simple implementation of the linear response protocol described above in which we need to study the linearized fluctuation of the field $\delta\tilde{\Phi}(r, v, \mathbf{k})$, dual to $\tilde{O}(t, \mathbf{k})$, about the gravitational solution (\ref{bmetric}) representing the dual nonequilibrium state. The initial condition for $\delta\tilde{\Phi}(r)$ at $v = v_{\rm in}$ is trivial by causality (since neither the pump nor the probe has been switched on) and the boundary condition is set by identifying the leading asymptotic $r\rightarrow 0$ mode with the probe source $\tilde{f}(t,\mathbf{k})$ i.e. imposing $\lim_{r\rightarrow 0}\delta\tilde{\Phi}(r, v, \mathbf{k})/r = \tilde{f}(v,\mathbf{k})$ for all times $v$. The solution for $\delta\tilde{\Phi}(r, v, \mathbf{k})$ is unique and can be obtained numerically via the method of characteristics \cite{Chesler:2008hg,Chesler:2013lia}. Further, we obtain the expectation value of the dual operator $\delta\langle\tilde{O}(t, \mathbf{k})\rangle$ from this solution via holographic renormalization \cite{Skenderis:2002wp}. Finally $G_R^{\tilde{O}\tilde{O}}(t, t',\mathbf{k})$ is extracted using the relation (\ref{causal}) with $\tilde{f}(v,\mathbf{k})$ chosen to be a narrow Gaussian profile appropriately normalized so that it can be treated as $\delta (v-t')$ up to any required order of numerical accuracy \cite{Banerjee:2016ray}. The latter feature then implies that $\delta\langle\tilde{O}(\mathbf{k})\rangle(t) = G_R^{\tilde{O}\tilde{O}}(t, t',\mathbf{k})$. More details are presented in the Appendix \ref{appen3}
. The above prescription reproduces the results obtained with the well-known Son-Starinets prescription \cite{Son:2002sd} for the thermal retarded correlation \cite{Banerjee:2016ray}. 

\section{\textbf{Exact results}}\label{exact} 
\begin{figure}[t]
\centering
\includegraphics[width=8cm,clip]{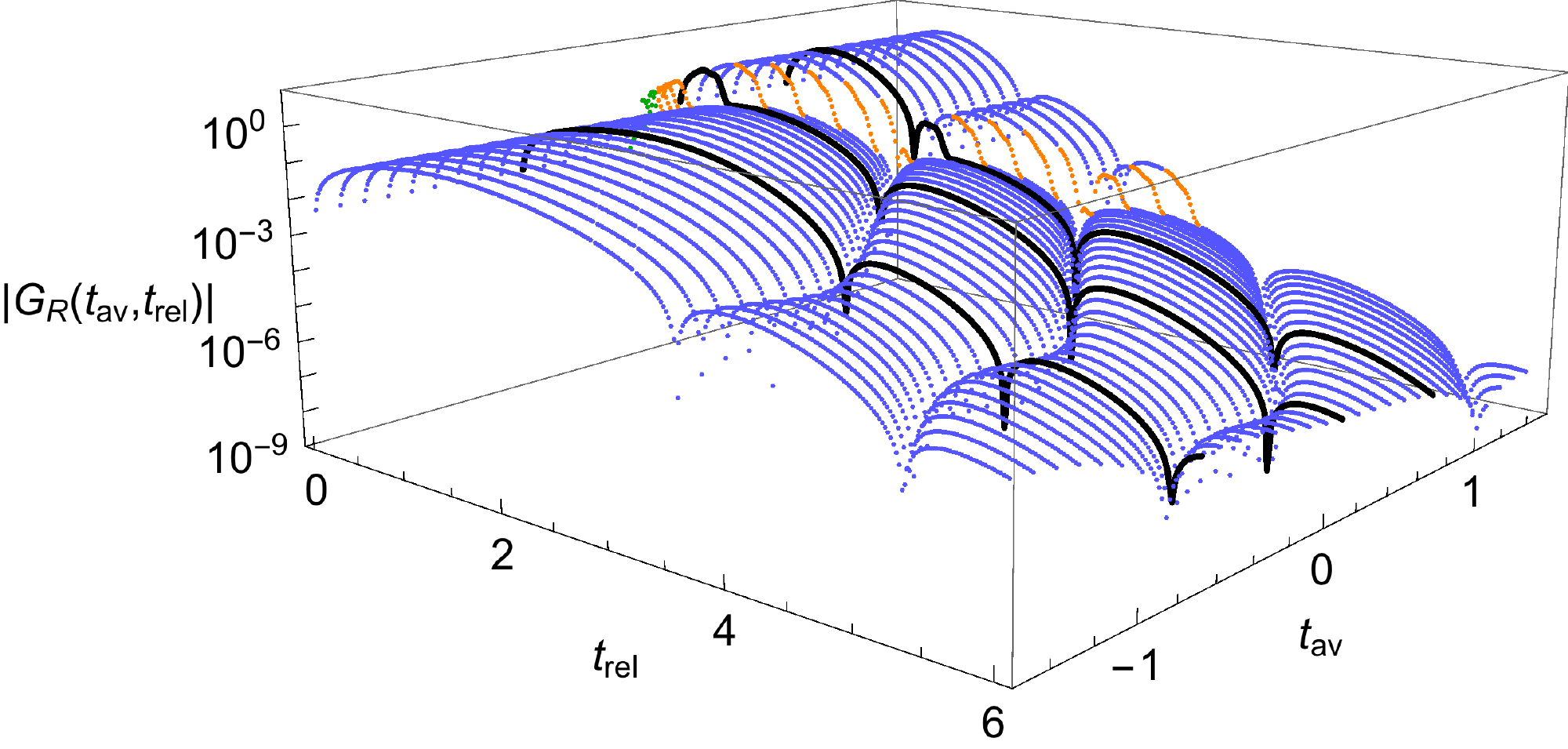}
\caption{The numerical result for $G_R$ as a function of $t_{\rm av}$ and $t_{\rm rel}$. Black curves indicate $G_R(t_{\rm rel})$ for constant representative values of $t_{\rm av}=-0.6$, $0.2$ and $0.8$ which are displayed in more detail in Fig. \ref{fig:cutgrid}. The $\mathcal{U}$, PP and POP regions (see text) are colored in blue, orange and green respectively.}
\label{fig:3DLogPlot}
\end{figure}

The exact results for $G_R^{\tilde{O}\tilde{O}}(t, t',\mathbf{k})$ in the driven nonequilibrium state are presented in Fig. \ref{fig:3DLogPlot}. We have set $\mathbf{k} = 0$ for presentation since we find (as in Ref. \cite{Banerjee:2016ray}) that our conclusions remain similar for $\vert\mathbf{k}\vert <T_{\rm in}$. For better understanding, instead of using the probe time $t'$ and observation time $t$ for the plot, we have used the average time $t_{\rm av} = (t+t')/2$ and the relative time $t_{\rm rel} = t-t'$. We have also subtracted a (state-independent) contact term from $G_R^{\tilde{O}\tilde{O}}(t_{\rm av}, t_{\rm rel})$ which is localized at $t_{\rm rel} = 0$. For the nonequilibrium geometry, we have chosen $\sigma = 0.03$ and $E_{\rm max} = T_{\rm in} = 1$ in Eq. \eqref{eq:pump}, which leads to the final temperature $T_{\rm f} \approx 2^{4/3}$.
Clearly for large values of $\vert t_{\rm av}\vert$, $G_R^{\tilde{O}\tilde{O}}(t_{\rm av}, t_{\rm rel})$ is independent of $t_{\rm av}$ and reduces to the initial/final thermal forms. Furthermore, since the underlying dynamics is conformal, at thermal equilibrium $G_R^{\tilde{O}\tilde{O}}(t_{\rm av},t_{\rm rel}) = T^2g_R^{\rm eq}(t_{\rm rel}T)$ (recall that $\Delta = 2$ for $\tilde{O}$). This is reflected in Fig. \ref{fig:3DLogPlot} through the appropriate $t_{\rm av}$ evolution in the width, slope and height of the \textit{ringdown} pattern of $G_R^{\tilde{O}\tilde{O}}(t_{\rm av}, t_{\rm rel})$ from its initial to final thermal forms. 

Remarkably, this $t_{\rm av}$ evolution of $G_R^{\tilde{O}\tilde{O}}(t_{\rm av}, t_{\rm rel})$ takes place over $-1.5 <  t_{\rm av} <1.5$ if we impose a cutoff on the maximum departures from the initial/final thermal forms by $10^{-4}$ times the respective maximal thermal values. This overall time scale ($ \approx 3$) is about $100$ times larger than the root-mean-square width $\sigma = 0.03$ of the pump and the time scale of evolution of one-point functions with the same cutoff. It is however commensurate with the time scale of the evolution of the location of the event horizon as we will see later. The association of the time scale of the departure from thermality of the retarded correlator to the time scale of event horizon evolution is an expected feature because the geodetic distance of the event horizon from the boundary controls the rate of dissipation, i.e. the effective quasinormal mode pole. These time scales can be made precise via wavelet analysis \cite{Mallat:2008:WTS:1525499} but will not be attempted here. It is worthwhile to note that since the event horizon responds acausally to the pump, $G_R^{\tilde{O}\tilde{O}}(t_{\rm av}, t_{\rm rel})$ starts evolving even when $t_{av} < -3 \sigma$, i.e. before the pumping is significant. However, the latter behavior is not really acausal for $G_R^{\tilde{O}\tilde{O}}(t_{\rm av}, t_{\rm rel})$ is nonlocal in time by definition and the causality merely implies it vanishes for $t_{rel} < 0$.

We divide the $t_{\rm av}-t_{\rm rel}$ plane into a \textit{universal} and a nonuniversal region. The universal $\mathcal{U}$ region is defined as the region where the probe time $t' = t_{\rm av} - t_{\rm rel}/2$ is away from the pumping duration, i.e. $\mathcal{U} :=  \{ t_{\rm rel}<2(t_{\rm av} -3\sigma)\}\cup \{ t_{\rm rel}>2(t_{\rm av} + 3\sigma)\}$. The features of $G_R(t_{\rm av}, t_{\rm rel})$ in this region can be attributed mostly to the change of the location of the event horizon which will turn out to be universal, i.e. \textit{independent of the details of the pumping protocol}, and determined largely by the initial and final temperatures, and the pumping duration only as we will see below. Therefore, these features are independent of the details of the gravity theory and hence the microscopic dynamics.

The nonuniversal region is further divided into two subregions (see Fig. \ref{fig:3DLogPlot}). The first is the \textit{probe-on-pump} (PP) region which is defined as where the probe time $t' = t_{\rm av} - t_{\rm rel}/2$ is \textit{within} but the observation time $t = t_{\rm av} + t_{\rm rel}/2$ is \textit{away from} the pumping time, i.e. $\vert t'\vert < 3\sigma$ and $\vert t\vert >3\sigma$. The features of $G_R(t_{\rm av},t_{\rm rel})$ in this region (most prominently an extra bump on the ringdown pattern) are not independent of the details of the microscopic dynamics. We will see below that these can be attributed nevertheless to the location of the apparent horizon (which responds to the pump instantaneously) and can be reproduced by a simple density matrix approximation dual to a simple prototype geometry. Conversely, the features of $G_R(t_{\rm av},t_{\rm rel})$ in the $\mathcal{U}$ and PP regions will allow us to \textit{locate the event and apparent horizons, within the numerical accuracy, of the dual geometry} (\ref{bmetric}) respectively, thus, providing us with some information of the dual gravity theory (i.e. the microscopic dynamics) as described below. The second (very tiny) subregion is the \textit{probe-and-observation-on-pump}  (POP) region where \textit{both} the probe and observer times are within the pumping time (i.e. $\vert t \vert, \vert t' \vert < 3\sigma$). The features here cannot be reproduced by our simple approximations and hence will not be analyzed here.
\begin{figure*}[t]
\centering
\includegraphics[width=\textwidth,clip]{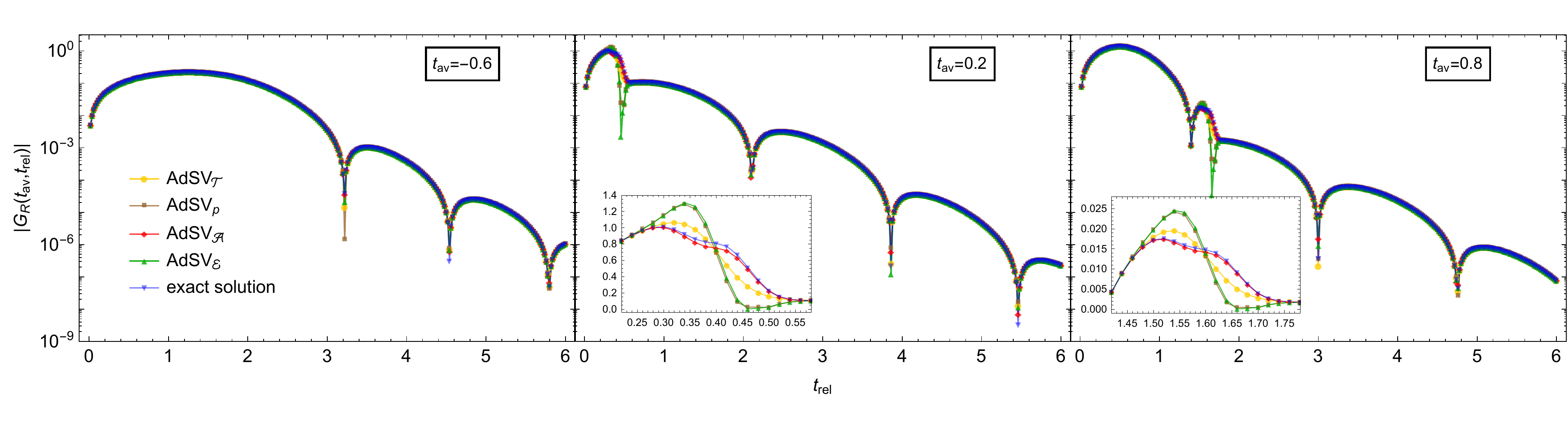}
\caption{Comparison of the exact retarded correlation to that obtained from the different AdSV geometries for three values of $t_{\rm av}$ -- the insets display zoom-ins on the PP region on a \textit{linear} scale.}
\label{fig:cutgrid}      
\end{figure*}

\section{\textbf{Simple density matrices and dual geometric prototypes}}\label{sec:AdSV}
 The trial nonequilibrium geometries are AdSV geometries defined only by a time-dependent black hole mass function $M(v)$ such that the functions $A(r,v)$ and $S(r,v)$ in Eq. (\ref{bmetric}) take the radically simple forms:
\begin{equation}\label{AdSV}
A(r,v) = 1 - M(v)r^3, \quad S(r,v) = 1/r,
\end{equation}
while the scalar field $\Phi$ vanishes. These AdSV geometries are not viewed here as solutions of the dual classical gravity theory but rather as simple prototypes for the actual numerical background. Using these simple geometric prototypes is analogous to representing the exact density matrices by instantaneously thermal density matrices 
\begin{equation}
\rho_{\rm inst}[T(t)]:=\exp(-H_{\rm CFT}/T(t))/{\rm Tr}(\exp(-H_{\rm CFT}/T(t)))
\end{equation}
which are not the solutions of the microscopic time-evolution equations either. In fact this identification is natural because neither in the density matrix nor in the AdSV geometries, relaxation (quasinormal) modes are excited. In the AdSV geometry the black hole changes its Hawking temperature which therefore is identified with the time-dependent temperature $T(t)$ in the density matrix. Thus $T(t)$ can obtained from $M(v)$ via
\begin{equation}
M(t) =  (4\pi/3) l^2 T^3(t).
\end{equation}
The term \textit{instantaneously thermal} as a qualifier for the density is understood in the above sense as being defined by an instantaneous temperature. However, if $T(t)$ changes fast so that $({\rm d}T(t)/{\rm d}t)\times (t_{\rm s}/ T(t)) \gg 1$ with $t_{\rm s}$ being the average scattering time, a typical observable will not be able to  adjust itself to its instantaneous thermal value although the density matrix can be described by an instantaneous temperature.

The simplest choice of $M(v)$ can be defined as a monotonic interpolation between the initial and final black hole masses ($M_{\rm in}$ and $M_{\rm f}$ respectively) that is readily provided by a tanh function whose time scale of variation is the same as the root-mean-square width $\sigma$ of the pump, i.e.
\begin{equation}\label{tanh}
M(v) = M_{\rm in} + ((M_{\rm f} -M_{\rm in})/2)\left(1 + \tanh\left(v/\sigma\right)\right).
\end{equation}
The above choice defines the $AdSV_{\mathcal{T}}$ prototype geometry examined in Ref. \cite{Banerjee:2016ray}. Crucially this prototype geometric approximation is parametrized \textit{just by the initial and final temperatures and the pump duration}, and thus can be constructed without prior knowledge of the dual gravity theory. 

The other prototype geometries important for the present discussion are $AdSV_{\mathcal{A}}$  and $AdSV_{\mathcal{E}}$ which can be constructed such that they reproduce the exact locations $r_{\rm EH}(v)$ and $r_{\rm AH}(v)$ of the event and apparent horizons of the numerical geometry respectively (see Appendix \ref{appen4}
for details and an explanation why ambiguities arising from diffeomorphism symmetries can be avoided). Finally, we also consider $AdSV_{p}$ as suggested in Ref. \cite{Bhattacharyya:2009uu} where $M(v)$ is designed to reproduce  $a_3(v)=\lim_{r\rightarrow0}(1/6)\partial_r^3A(r,v)$ of the numerical geometry exactly.  Holographic renormalization tells us that $P(v) = \langle t_{xx}(v)\rangle  = \langle t_{yy}(v) \rangle =-a_3(v)$ is the pressure, and therefore $AdSV_{p}$ reproduces the exact time evolution of the pressure by construction. 
\begin{figure}[h]
\centering
\includegraphics[width=8cm,clip]{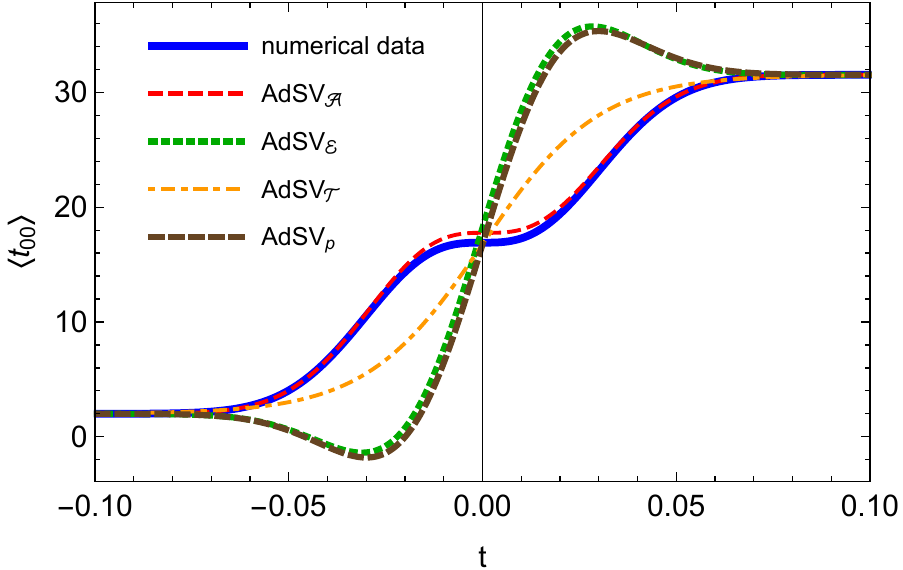}
\caption{Plots comparing the exact $\langle t_{00}(t)\rangle$ with those obtained from the prototype AdSV geometries.}
\label{fig:energyden}
\end{figure}

For small amplitude pumps, the AdSV geometries can be good approximations to the exact numerical geometry dual to nonequilibrium states \cite{Bhattacharyya:2009uu}, whereas in the arbitrary and large-amplitude regime (which produces a large difference $\Delta T$ between final and initial temperatures compared to the inverse of time period of energy injection), which we have considered in this work, we find that none of the AdSV approximations can reproduce \textit{both} the time dependence of the pressure and energy density accurately. We have demonstrated this observation in Fig. \ref{fig:energyden}. Note that the prototype $AdSV_{\mathcal{A}}$ geometry gives $\langle t_{00}(t)\rangle$ very accurately. Observe that the curves for $\langle t_{00}(t)\rangle$ of $AdSV_{\mathcal{E}}$ and $AdSV_{p}$ geometries closely follow each other. For the AdSV geometries, $\langle t_{00}(t)\rangle = 2\langle t_{xx}(t)\rangle$. Since $AdSV_{p}$ reproduces the numerical pressure $\langle t_{xx}(t)\rangle$ exactly, $AdSV_{\mathcal{E}}$ provides a very good approximation to the pressure. Therefore, the AdSV prototype approximations demonstrate that the event and apparent horizons of the dual geometry carry information about the time evolution of the pressure and the energy-density respectively.

The comparisons of the time-dependent retarded correlation of the prototype geometries with that of the exact one are presented in Fig. \ref{fig:cutgrid} where we have chosen three representative values of $t_{\rm av}$ which are $-0.6$, $0.2$ and $0.8$. The insets for $t_{\rm av} = 0.2$ and $0.8$ show the PP regions on a linear scale. 

\section{\textbf{Universal features, effective description and the horizons}}\label{horizons} 
\begin{figure}[t]
\centering
\includegraphics[width=8cm]{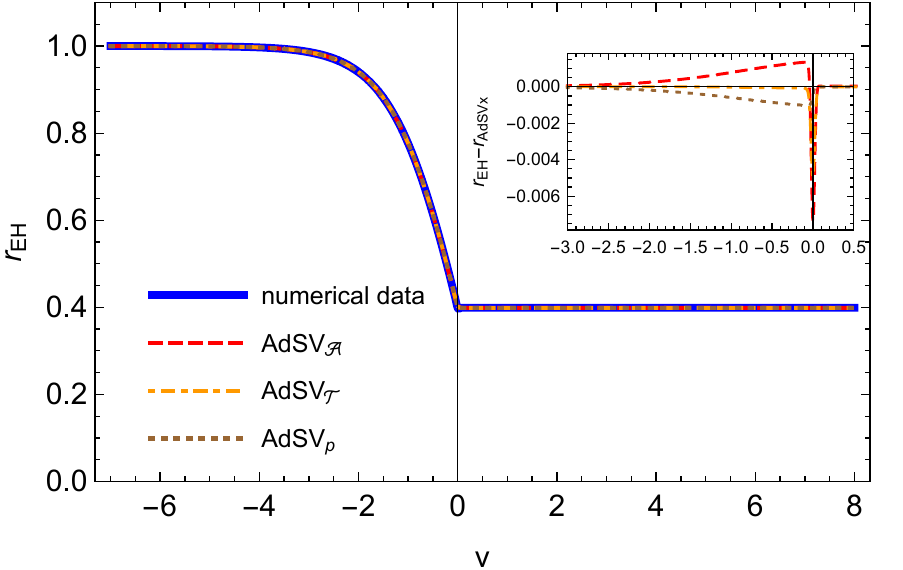}
\caption{The locations of the event horizon $r_{\rm EH}$ in the different AdSVs and the exact numerical value with the inset focusing on their differences.}
\label{fig:EHLocs}
\end{figure}

We readily see from Fig. \ref{fig:cutgrid} that in the $\mathcal{U}$ region, all AdSV geometries approximate the exact retarded propagator within $1$ percent relative accuracy which is much better than a relative factor of $\mathcal{O}(3\sigma T_{\rm in})$ i.e. $10$ percent accuracy that can be taken as a benchmark \footnote{To obtain a naive estimate of error produced by the AdSV approximation, one may note that it ignores the effect of the gravitational quasinormal mode (QNM). Furthermore, the AdSV approximations will deviate from the exact background geometries appreciably only within the half-width of the pumping duration. Therefore it can be expected that the AdSV approximations will deviate from the exact geometry by the QNM attenuation factor of $1- \exp(-3\sigma T_{\rm in}) \approx 3\sigma T_{\rm in}$. This is however not a rigorous estimate because it ignores the effect of nonlinearities of gravitational dynamics. In any case, this gives a benchmark for the accuracy expected from an AdSV approximation and we find that the AdSV prototypes discussed perform better than this expectation.}. A detailed analysis of the standard deviation as shown in Fig. \ref{fig:NDplot} (for details see Appendix \ref{appen5}) reveals that the best approximation in the $\mathcal{U}$ region is provided by the $AdSV_{\mathcal{E}}$ prototype that reproduces the exact location of the event horizon. Remarkably, the approximations of $G_R(t_{\rm av}, t_{\rm rel})$ provided by the prototype geometries in the universal $\mathcal{U}$ region can be intuitively explained by the fact that all these prototype geometries including $AdSV_{\mathcal{T}}$, $AdSV_{\mathcal{A}}$ and $AdSV_{p}$ also reproduce the exact locations of the time-dependent event horizon within $1$ percent relative accuracy for all times as shown in Fig. \ref{fig:EHLocs}. In particular the $AdSV_{\mathcal{T}}$ geometry which can be constructed without any detailed knowledge of the dual gravity theory reproduces the features of $G_R(t_{\rm av}, t_{\rm rel})$ in the $\mathcal{U}$ region and also the location of the event horizon with remarkably high accuracy even \textit{within} the pumping duration (see the inset plot in Fig. \ref{fig:EHLocs}). So we can lay claim to their universality as mentioned before. The $AdSV_{\mathcal{T}}$ geometries have the feature that the location of the event horizon is determined mostly by the difference of the final and initial temperatures $\Delta T$ and is independent of the pumping duration ($6\sigma$) provided $3\sigma\Delta T \ll 1$ \footnote{See Fig. (1c) in Ref. \cite{Banerjee:2016ray}.}. This can be arranged in the numerical geometry if $E_{\rm max}\geq T_{\rm in}$ and $\sigma T_{\rm in}  \ll 1$, and therefore under these conditions the features in the $\mathcal{U}$ region should also be universal.

Conversely, since $AdSV_{\mathcal{E}}$ gives the best approximation in the $\mathcal{U}$ region, we conjecture that the event horizon in the geometry (\ref{bmetric}) dual to the driven nonequilibrium state can be located by finding that prototype AdSV geometry fitting the universal features of $G_R(t_{\rm av}, t_{\rm rel})$ best.

The inset plots in Fig. \ref{fig:cutgrid} clearly show that the prototype $AdSV_{\mathcal{A}}$ geometry which reproduces the exact location of the apparent horizon in bulk coordinates (\ref{bmetric}) provides a remarkably good approximation to $G_R(t_{\rm av}, t_{\rm rel})$ in the PP region where the pump protocol plays a dominant role with about $1$ percent relative accuracy. It is remarkable that $AdSV_{\mathcal{T}}$ also provides a reasonably good accuracy in the PP region while $AdSV_{\mathcal{E}}$ and $AdSV_{p}$ results have large standard deviations from the exact results as clearly visible in Fig. \ref{fig:NDplot}. We can also check that $AdSV_{\mathcal{T}}$ provides a reasonable approximation to the exact location of the apparent horizon while $AdSV_{\mathcal{E}}$ and $AdSV_p$ fail to do so. 

From these results it follows that an effective theory for $G_R(t_{\rm av}, t_{\rm rel})$ including the $\mathcal{U}$ and PP regions (but excluding a tiny POP region) can be obtained simply by knowing how the pump protocol determines the apparent horizon in the dual geometry and then using the simple prototype $AdSV_{\mathcal{A}}$ geometry to reproduce $G_R(t_{\rm av}, t_{\rm rel})$. Our results also provide sufficient support for our conjecture that the location of the apparent horizon in the bulk coordinates (\ref{bmetric}) can be deciphered from $G_R(t_{\rm av}, t_{\rm rel})$ simply by finding which prototype AdSV geometry provides the best approximation in the PP region. We would like to point out that even though the above conjecture has been put forth based on the observation derived from a scalar field coupled to Einstein gravity, we believe that our claim regarding the reconstruction of the horizons should hold for general two-derivative
gravity theories (eg. Einstein-Maxwell-dilaton type theories). We leave the numerical check of the same for future works.

To be able to approximate the retarded correlator of an exact numerical geometry with the retarded correlator obtained from AdSV geometries is a highly nontrivial result. For example, if we consider a one-point function such as the time-dependent pressure $P(t)$ in a homogeneous quench, we can always design an AdSV geometry which has the right time-dependent mass $M(v)$ of the black hole such that it reproduces $P(t)$ exactly. However, the nonequilibrium retarded correlator $G_R(t; t')$ depends on both $t$ and $t'$, since the background is not time-translation invariant. Therefore, it would be rather impossible to fit $G_R(t; t')$, specially in the PP domain, just by choosing $M(v)$ which is a function of one (and not two) variables. The claim we make becomes significant in this regard. It is that the AdSV which gives the best approximation to the $G_R(t; t')$ in the PP $(\mathcal{U})$ domain should also reproduce the location of the apparent (event) horizon, within numerical precision, of the actual dual geometry. In fact, this allows us to use the prototype AdSV approximations to learn something about dual bulk geometries without knowing the dual theory of gravity.

Finally, as is clear from Fig. \ref{fig:cutgrid}, the prototype $AdSV_{\mathcal{T}}$ geometry provides a reasonably good approximation for $G_R(t_{\rm av}, t_{\rm rel})$ in the entire $t_{\rm av}-t_{\rm rel}$ plane (except for the tiny POP region) and therefore, someone interested in finding signatures of a large-$N$ conformal strong coupling regime in $G_R(t_{\rm av}, t_{\rm rel})$ can readily utilize this $AdSV_{\mathcal{T}}$ geometry constructed from simple measurable inputs.

\section{\textbf{Concluding remarks}}\label{Conclusions} Our results strongly indicate that nonequilibrium behavior of correlation functions should play a crucial role in both applications and the fundamental understanding of the holographic principle. From the perspective of applications, we have found remarkable universal features of the time-dependent causal correlations that can be reproduced from simple prototype geometric approximations without the detailed knowledge of the dual gravity theory (i.e. the microscopic dynamics). From the perspective of fundamental understanding, we obtained interesting pointers regarding how we can construct the dual gravitational theory directly from appropriate observables. It is to be noted that some of our results parallel interesting universal behavior of equal-time correlations and the entanglement entropy during fast quenches \cite{Das:2014lda,Das:2014hqa}.
\begin{figure}[t]
\centering
\includegraphics[width=8cm]{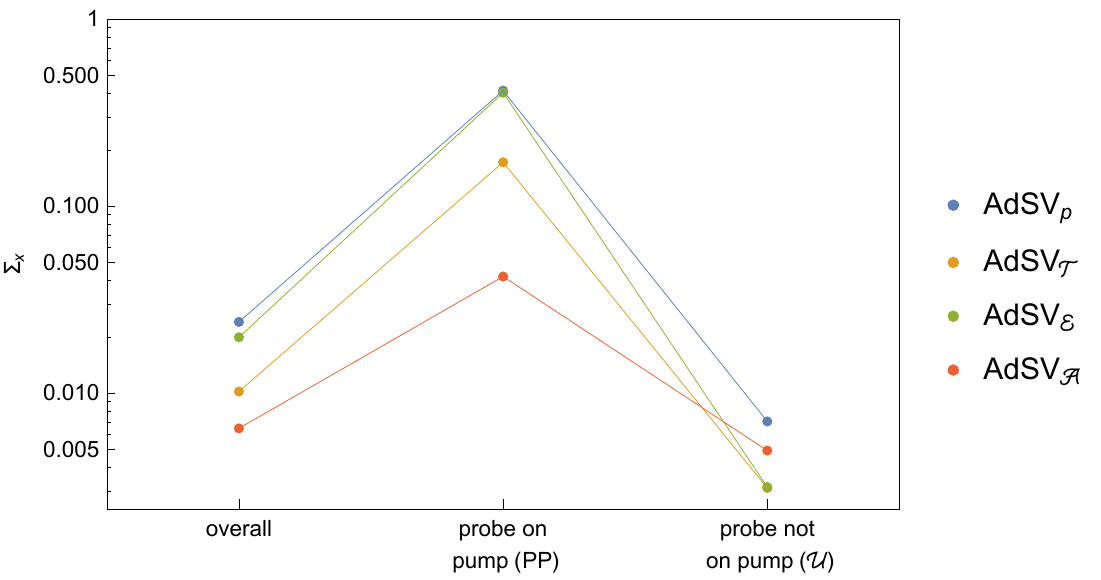}
\caption{Plot of standard deviations $\Sigma_x$ (defined in \eqref{eq:measureApprox1}) of various $AdSV_{x}$ approximations from calculated values of $G_R(t, t')$ obtained from the numerical geometry in three categories: (i) overall, i.e. including all values of $t$ and $t'$, (ii) in the PP (probe on pump) region and (iii) the universal $\mathcal{U}$ region.}
\label{fig:NDplot}
\end{figure}
\begin{acknowledgments}
The research of A.M. is supported by a Lise-Meitner fellowship of the Austrian Science Fund (FWF), project no. M 1893-N27. We thank S. Banerjee and T. Ishii for collaboration during the early stages of this project, and C. Ecker for helpful discussions. We are very grateful to A. Rebhan and A. Soloviev for carefully reading and discussing the draft.
\end{acknowledgments}

\begin{appendix}
\section{\textbf{Numerical calculation of the geometry dual to the driven nonequilibrium state using the method of characteristics}}
\label{appen1} 
The $(3+1)$-dimensional gravitational equations are
\begin{eqnarray*}\label{baction}
R_{MN} + \frac{3}{l^2} G_{MN} &=& T_{MN}[\Phi] - \frac{1}{2}G_{MN} G^{PQ}T_{PQ}[\Phi],\\
(\Box - m^2)\Phi &=& 0,
\end{eqnarray*}
where $\Box$ is the Laplacian operator constructed from the metric $G_{MN}$ and
\begin{equation*}\label{st}
T_{MN}[\Phi] = \frac{1}{2}\partial_M \Phi\partial_N \Phi -\frac{1}{4}G_{MN}G^{PQ}\partial_P \Phi\partial_Q \Phi - \frac{1}{2}m^2\Phi^2.
\end{equation*}
We choose $m^2=-2/l^2$ as described before. For the sake of convenience, we can set $l=1$. 

Denoting the derivative along the outgoing null direction as $d_+=\partial_v- A / 2~ \partial_r$, the equations of motion for the scalar field and the metric [i.e. those of $A(r,v)$, $S(r,v)$ and $\Phi(r,v)$ appearing in Eq. (2)]
) are as follows:
\begin{subequations}
\begin{align}
\partial_r^2S +\frac{2}{r}\partial_rS +\frac{(\partial_r\Phi)^2}{4}S&=0\label{eqS}~,\\
\partial_r d_+S+ \frac{\partial_rS d_+S}{S}+\frac{S}{4r^2}\left(6-\frac{m^2}{2}\Phi^2\right)&=0\label{eqdpS}~,\\
r \partial_r^2\frac{A}{r} +\frac{4 \partial_rS d_+ S}{S^2}-\partial_r\Phi d_+\Phi &=0~,\label{eqA}\\
d_+^2S+\frac{r^2}{2}\partial_r \frac{A}{r^2} d_+S +\frac{S(d_+\Phi)^2}{4}&=0\label{eqdpdpS}~,\\
\partial_r(d_+\Phi)+\frac{\partial_rS d_+\Phi}{S}+\frac{\partial_r\Phi d_+S}{S}+\frac{m^2}{2r^2}\Phi&=0\label{eqphi}~.
\end{align}
\end{subequations}
As described before, the above equations have unique solutions with specified initial and boundary conditions which have been chosen by (i) imposing that in the far past $v_{\rm in} \ll -3\sigma$ the geometry is a static AdS black brane (planar black hole) with a vanishing $\Phi$ field and (ii) the dual system lives in flat Minkowski space and that the energy is pumped through a source $f(v)  = \lim_{r\rightarrow 0}\Phi(r,v)/r$ coupling to the operator $O$ dual to the scalar field $\Phi$. The numerical protocol involves the following steps: 
\begin{enumerate}
\item{At initial time $v = v_{\rm in}$, we solve Eq. \eqref{eqS} by radial integration to obtain $S(r,v_{\rm in})$ uniquely using the asymptotic expansion 
\begin{equation}\label{Sasymp}
S(r,v) = \frac{1}{r} - \frac{f^2(v)}{8} r + \mathcal{O}(r^2)
\end{equation}
(obtained from the radial expansion of the equations of motion) which holds for all $v$ and the initial condition $\Phi (r, v_{\rm in}) = 0$. }
\item{Next we solve Eq. \eqref{eqdpS} to obtain $d_+S(r, v_{\rm in})$ by radial integration using the boundary condition $d_+S(r= 0,v) \approx 1/(2r^2)$ (which holds for all $v$), the initial condition $\Phi (r, v_{\rm in}) = 0$ and the solution for $S(r, v_{\rm in})$.}
\item{With the knowledge of $S, d_+S $ and $\Phi$, we solve Eq. \eqref{eqphi} to obtain $d_+\Phi (r,v_{\rm in})$ using the boundary condition $d_+\Phi(r = 0,v) = - f(v)/2 $ (which holds for all $v$).}
\item{$A$ can then be found uniquely from Eq. \eqref{eqA} at $v = v_{\rm in}$ using the asymptotic expansion 
\begin{equation}\label{Aasymp}
A(r,v) = 1 - \frac{f^2(v)}{4} r^2 + a_3(v) r^3 +\mathcal{O}(r^4)
\end{equation}
(obtained from the radial expansion of the equations of motion) which holds for all $v$. At $v= v_{\rm in}$, we need to input $a_3 = -M_{\rm in}$.}
\item{From the definition of $d_+$ it follows that $\partial_v\Phi =d_+\Phi +(A/2) \partial_r\Phi$ using which we find $\partial_v \Phi$ at initial time $v_{\rm in}$ since $A$ and $d_+\Phi$ have been obtained in the previous steps. By using a time stepper we then step up to next time $v_{\rm in}+\Delta v$ to obtain $\Phi(v_{\rm in}+\Delta v)$.}
\item{The equation \eqref{eqdpdpS} is actually a constraint, therefore if it is satisfied at $r=0$ then it should be satisfied  for all $r$. The leading non-trivial asymptotic term of this equation yields the time-evolution of $a_3(v)$ which takes the form:
\begin{equation}\label{eom-M}
\partial_v a_3= \frac{1}{2}f(v)(\partial_v^2 f(v)-\partial_v f_1(v)),
\end{equation}
where $f_1(v) =\lim_{r\rightarrow0}(1/2)\partial_r^2\Phi(r,v)$. This equation reproduces the CFT Ward identity corresponding to energy conservation (see below). We can use this to update the value of $a_3$ to obtain $a_3(v_{\rm in}+\Delta v)$. }
\item{Having known $\Phi(r,v_{\rm in}+\Delta v)$ and $a_3(v_{\rm in}+\Delta v)$ at the next time step,  we start again from step 1 to solve for all the other functions at this time step. We repeat time steps until we reach the final black brane geometry with a vanishing (rather sufficiently small) $\Phi$ field.}
\end{enumerate}
For the radial integration we have used a pseudospectral method with 30 grid points and for stepping up in time we have used the Adams--Bashforth fourth-order time stepper with a time step of $\delta v=.0003$ (which is $0.01\sigma$). A suitable numerical domain $0 < r < r_c$ has been chosen so that the apparent horizon and hence the event horizon of the geometry lies within this domain.

\section{\textbf{Details of holographic renormalization}} \label{appen2}
We follow the minimal subtraction scheme to obtain
\begin{subequations}
\begin{align}
\langle t_{00}\rangle &=\epsilon = \mathcal{C}(-2 a_3  +f (\partial_v f - f_1)), \\
\langle t_{xx}\rangle &= \langle t_{yy}\rangle =P =-  \mathcal{C}a_3,\\  
\langle O\rangle &=   \mathcal{C}(f_1-\partial_v f), \label{Oren} 
\end{align} 
\end{subequations}
where $a_3 = \lim_{r\rightarrow0}\partial_r^3A(r,v)/6$, $f_1 =  \lim_{r\rightarrow0}\partial_r^2\Phi(r,v)/2$ and $\mathcal{C}$ is an overall factor proportional to $N^2$ of the dual theory. We readily see that the above along with Eq. \eqref{eom-M} imply the CFT Ward identities
\begin{subequations}
\begin{align}
\partial_t \langle t^{0}_{\phantom{\mu}0}\rangle &=-\partial_t\epsilon \,= \langle O\rangle\partial_t J, \\\langle{\rm Tr} t\rangle &= 2P -\epsilon = (d -\Delta)J\langle O\rangle
\end{align} 
\end{subequations}
identifying $v$ with the field-theory time coordinate $t$ at the boundary $r= 0$ and $J$ with $f$ as mentioned before, and using $\Delta =2$ and $d=3$. In the AdSV geometries, there is no scalar field and hence $\langle O\rangle = 0$, $\langle t_{00}\rangle = -2\mathcal{C}a_3(v)$ and $ \langle t_{xx}\rangle = \langle t_{yy}\rangle = -\mathcal{C}a_3(v)$. We normalize the definitions of all one-point functions by dividing them by a suitable factor proportional to $N^2$ so that we can set $\mathcal{C} = 1$.

Note that for the probe scalar field $\tilde{\Phi}$, the same formula \eqref{Oren} applies to the dual $\langle\tilde{O}\rangle$ operator both in the numerical and the AdSV geometries with $f$ and $f_1$ replaced by $\tilde{f} = \lim_{r\rightarrow 0}\tilde{\Phi}(r,v)/r$ and $\tilde{f}_1= \lim_{r\rightarrow 0}\tilde{\Phi}(r,v)/r^2$ respectively since $\langle\tilde{O}\rangle$ has the same scaling dimension $\Delta  = 2$. It is easy to see that the contact term $-\partial_v \tilde{f} = -\partial_t \tilde{J}$ in $\langle\tilde{O}\rangle$ gives a $- \partial_{t_{\rm rel}}\delta(t_{\rm rel})$ contribution to  $G_R(t_{\rm av},t_{\rm rel})$. After using the Wigner transform, i.e. the Fourier transform of $t_{\rm rel}$ dependence (defined according to our sign convention), we obtain 
\begin{equation}
G_R(t_{\rm av},\omega) = \int_{-\infty}^\infty e^{-i\omega t_{\rm rel}}G_R(t_{\rm av},t_{\rm rel}) \mathrm{d}t_{\rm rel}.
\end{equation} 
The contact term $- \partial_{t_{\rm rel}}\delta(t_{\rm rel})$ then produces $-i\omega$ term in $G_R(t_{\rm av},\omega)$. This implies that the spectral function 
\begin{equation}
\rho(t_{\rm av}, \omega) = - 2{\rm Im}G_R(t_{\rm av}, \omega)
\end{equation}
 gets the state-independent contribution $2\omega$ and we have checked  that this term ensures that $\rho(\omega, t_{\rm av}) > 0$ for $\omega >0$ as should follow from the spectral representation in the dual field theory. 

\section{\textbf{Numerical calculation of the nonequilibrium retarded correlation function}}
\label{appen3}
 The Klein-Gordon equation for the scalar field $\tilde{\Phi}(r, v, \mathbf{k})$ dual to the operator $\tilde{O}$ whose correlation function we are calculating takes the same form as Eq. \eqref{eqphi} for $\mathbf{k} = 0$ with $S(r,v)$ and $A(r,v)$ fixed by their forms in the background geometry dual to the driven nonequilibrium state. The method of characteristics can then be readily implemented provided we specify $\tilde{\Phi}(r,v)$ at an initial time and the source $\tilde{f}(v) = \lim_{r\rightarrow0}\tilde{\Phi}(r,v)/r$ for all $v$ as should be clear from our prior discussion. In the AdSV backgrounds, $S(r,v) = 1/r$ and $A(r,v) = 1- M(v)r^3$. In order to obtain $G_R^{\tilde{O}\tilde{O}}(t, t')$, we need to set initial conditions $\tilde{\Phi}(r,v_{\rm in}) = 0$ and $\tilde{f}(v) = \delta (v-t')$ as discussed in the main text. The choice $\tilde{f}(v) = \delta (v-t')$ and Eq. \eqref{Oren} simply imply that $G_R^{\tilde{O}\tilde{O}}(t, t') = \tilde{f}_1(t) - \partial_t\delta(t-t')$.

\begin{figure}[h!]
\includegraphics[scale=0.6]{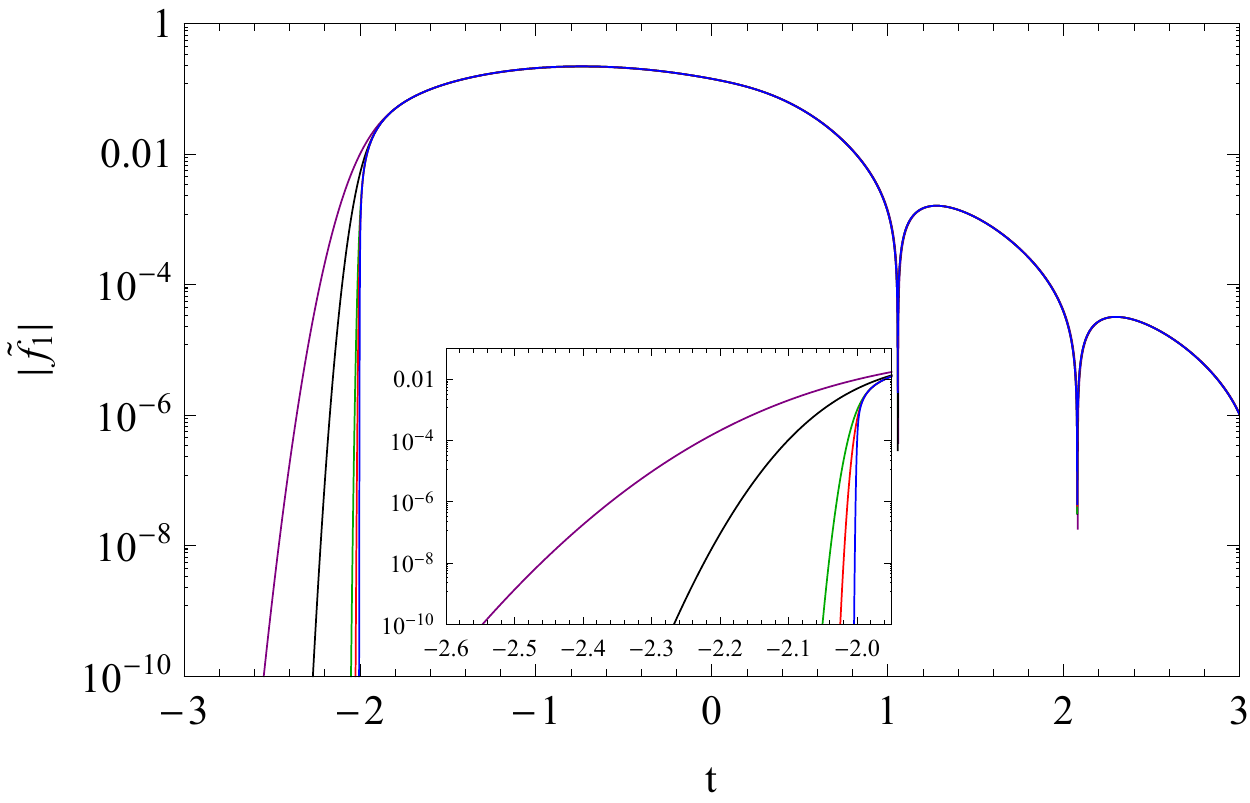}
\caption{For the Gaussian source $\tilde{f}$ centered at $t'=-2$, $|\tilde{f}_1|$ is plotted on a log scale for various values of $\tilde{\sigma}$. The purple, black, green, red and blue colours correspond to the widths $\tilde{\sigma}= 0.1$, $0.05$, $0.01$, $0.005$, and $0.001$ respectively.}
\label{fig:diracdelta}
\end{figure}

We now show that numerically the delta function limit for the source $\tilde{f}(v)$ can be taken by assuming it to be a normalized Gaussian function
\begin{equation}
\tilde{f}(v)= \frac{1}{\sqrt{2\pi} \tilde{\sigma}} e^{\frac{(v-t')^2}{2\tilde{\sigma}^{2}}},
\end{equation} 
with root-mean-square width $\tilde{\sigma}$ [not to be confused with the $\sigma$ of the pump $f(t)$] and taking the limit $\tilde{\sigma} \rightarrow 0$. We have plotted $|\tilde{f}_1(t)|$ in Fig. \ref{fig:diracdelta} for various values of $\tilde{\sigma}$. We readily see that the behavior of $\tilde{f}_1(t)$ converges with decreasing $\tilde{\sigma}$ for \textit{all} times $t$ up to any desired order to numerical accuracy. For practical purposes, we choose to work with the width $\tilde{\sigma}=.01$ which allows to take the Dirac delta limit at the level of four-digit precision. 

\section{\textbf{Construction of the prototype geometries $AdSV_{\mathcal{A}}$  and $AdSV_{\mathcal{E}}$}} \label{appen4} We first address the issue of how we can compare the numerical geometry (2) 
representing the dual of the driven nonequilibrium state with the prototype AdSV geometries (5) 
representing simple density matrices given that they are different spacetimes. First, of course for all these geometries we are choosing the ingoing Eddington-Finkelstein gauge where 
\begin{equation}\label{EF}
G_{rr}= G_{rx} = G_{ry} =0, \qquad G_{rv} = -l^2/r^2.
\end{equation}
Second, all of these geometries coincide at very early and very late times corresponding to AdS black branes (planar black holes) with masses $M_{\rm in}$ and $M_{\rm f}$ respectively. This however does not completely fix the choice of coordinates in these geometries exactly because the gauge \eqref{EF} has  residual diffeomorphism symmetries corresponding to 
\begin{equation}
r \rightarrow r + \lambda(v), \quad  v \rightarrow v + v_0
\end{equation}
with $v_0$ being a constant. Note that actually if we replace $\lambda(v)$ in the radial diffeomorphism above by $\lambda(v,x,y)$, the gauge \eqref{EF} is still preserved but the metric is no longer manifestly homogeneous. Also note that under this (radial) diffeomorphism, the functions $A$ and $S$ transform to $\tilde{A}$ and $\tilde{S}$ respectively which have different asymptotic behaviors at $r=0$. Therefore this diffeomorphism symmetry is simply fixed by our boundary conditions $\lim_{r\rightarrow0}A= \lim_{r\rightarrow0}S = 1$. The second residual diffeomorphism symmetry corresponding to time translation is fixed in the numerical geometry (2) 
by choosing the time in which the pump $f(t)$ peaks to be $t=0$ (recall that at $r=0$, the bulk time coordinate $v$ coincides with the field-theory time coordinate $t$). In the prototype AdSV geometries, this time translation freedom is fixed by a suitable form of matching with the numerical geometry. In $AdSV_{\mathcal{T}}$ where $M(v)$ is chosen to be as shown in Eq. (6) 
, the origin of time is chosen by demanding that the midpoint of the $\tanh$ function coincides with the time when the pumping is peaked at the boundary. In $AdSV_p$, this is fixed by construction via the exact matching of $a_3(v)$ [i.e. $P(t)$] with that obtained from the numerical geometry. Similarly, it will be fixed in the $AdSV_{\mathcal{A}}$ and $AdSV_{\mathcal{E}}$ by construction via the exact matching of the location of the apparent and the event horizons as described below.

The apparent horizon $r_{\rm{AH}}^{\rm exact}(v)$ of the numerical geometry (2)
 can be found by solving 
\begin{equation}
d_+S(v,r_{\rm{AH}}^{\rm exact}(v))=0 
\end{equation}
since $r = r^{\rm exact}_{\rm{AH}}(v)$ is a surface of vanishing extrinsic curvature. In the prototype AdSV geometries (5)
, this equation simplifies because $S(r,v) = 1$ and $A(r,v)$ takes a simple form determined by the choice of the mass function $M(v)$ so that the location of the apparent horizons are given simply by ${r_{\rm AH}^{\rm AdSV}}^3(v) = M(v)$. Therefore, in order to construct a prototype $AdSV_{\mathcal{A}}$ geometry which can reproduce the location of the apparent horizon of the actual numerical background $r_{\rm{AH}}^{\rm exact}(v)$, we need to choose the black hole mass function $M(v)$ as 
\begin{equation}
M(v)=\frac{1}{{r_{\rm AH}^{\rm exact}}^3(v)}.
\end{equation}
This completes the construction of $AdSV_{\mathcal{A}}$.

The location of the event horizon $r_{\rm{EH}}(v)$ is  given by the null geodesic which coincides with the location of the final black hole horizon in the limit $v\rightarrow\infty$ i.e. in the far future. For the numerical geometry (2)
, the location of the event horizon can be found by solving the differential equation
\begin{equation}
\partial_v r_{\rm{EH}}^{\rm exact}(v)+A(v,r_{\rm{EH}}^{\rm exact}(v))/2=0
\end{equation}
subject to the future boundary condition of the equilibrium horizon, i.e, $r_{\rm{EH}}^{\rm exact}(v\rightarrow \infty)=M_{\rm f}^{1/3}$. In the AdSV prototype geometries (5)
, the location of the event horizon $r_{\rm{EH}}^{\rm AdSV}(v)$ is given by the same equation but with $A(r,v)$ taking the simpler form determined by the choice of the mass function $M(v)$. It follows that in order to construct a prototype geometry $AdSV_{\mathcal E}$  whose event horizon coincides with that of the actual numerical geometry, the mass function $M(v)$ should be chosen to satisfy
\begin{equation}
M(v)=\frac{1}{{r_{\rm EH}^{\rm exact}}^3(v)}\left( 1+2 \partial_v r_{\rm EH}^{\rm exact}(v) \right).
\end{equation}
This completes the construction of $AdSV_{\mathcal{E}}$.
\section{\textbf{Quantitative comparison of $G_R(t_{\rm av}, t_{\rm rel})$ obtained in the prototype geometries}}
\label{appen5}
In order to quantify how well the prototype geometries approximate the exact $G_R(t_{\rm av}, t_{\rm rel})$ we need to compute
\begin{equation}\label{eq:measureApprox1}
\Sigma_x=\frac{\int_{\mathcal{R}} {\rm d}t_{\rm av}{\rm d}t_{\rm rel}\frac{\vert G_R(t_{\rm av}, t_{\rm rel})-G^{AdSV_{x}}_R(t_{\rm av}, t_{\rm rel})\vert}{\vert G_R(t_{\rm av}, t_{\rm rel})\vert}}{\int_{\mathcal{R}} {\rm d}t_{\rm av}{\rm d}t_{\rm rel}},
\end{equation}
where $x$ is $\mathcal{E}$, $\mathcal{A}$, $\mathcal{T}$ or $p$, and $\mathcal{R}$ denotes either the universal region $\mathcal{U}$ or the PP region. This we approximate by a Riemann sum over the plaquettes $\Delta t_{\rm av}\Delta t_{\rm rel}$. Furthermore, we introduce a cutoff by demanding $\vert G_R(t_{\rm av}, t_{\rm rel})\vert>10^{-4}$ (in units $T_{\rm in} = 1$) in both integrands of Eq. (22) 
 in order to tame numerical errors.

In the region $\mathcal{U}$ we find that all of the prototype geometries deviate from the exact result by less than 1 percent on average. In particular $AdSV_{\mathcal E}$ and $AdSV_{\mathcal T}$ perform equally well (according to our expectation because $AdSV_{\mathcal T}$ approximates the location of the event horizon very well) although $AdSV_{\mathcal E}$ gives a slightly better approximation. These are followed by $AdSV_{\mathcal A}$ while $AdSV_{p}$ performs the worst. In the PP region again $AdSV_{p}$ is the worst approximation while the ranking of the others is exactly reversed -- both $AdSV_{\mathcal A}$ followed by $AdSV_{\mathcal T}$ are better than 1 percent while both $AdSV_{\mathcal E}$ followed by $AdSV_{p}$ deviate by a bit more than 1 percent on average. We have also checked the POP region, where both the observation time as well as the probe time reside within the pump duration. However, none of the prototype geometries give reasonable results there (with the same ranking as in the PP region but deviations are up to 50 percent on average).

\end{appendix}

\bibliographystyle{apsrev4-1}
\bibliography{Paper2PRL}
\end{document}